\documentclass[12pt]{article}

\usepackage{graphicx}
\usepackage{color}

\makeatletter
\g@addto@macro\bfseries{\boldmath}
\makeatother

\def\jpsi{\hbox{$J\kern-0.15em/\kern-0.15em\psi\kern0.15em$}}

\def \beq{\begin{equation}}
\def \eeq{\end{equation}}
\def\bea{\begin{eqnarray}}
\def\eea{\end{eqnarray}}

\def\eqref#1{(\ref{#1})}

\def\URLtilde{\lower0.2em\hbox{$\tilde{\phantom{a}}$}}

\def\mycomm#1{\hfill\break\strut\kern-3em{\color{red}\tt ====> #1
\color{black}}\hfill\break}

\newcount\timecount
\newcount\hours \newcount\minutes  \newcount\temp \newcount\pmhours
\hours = \time
\divide\hours by 60
\temp = \hours
\multiply\temp by 60
\minutes = \time
\advance\minutes by -\temp
\def\hour{\the\hours}
\def\minute{\ifnum\minutes<10 0\the\minutes
\else\the\minutes\fi}
\def\clock{
\ifnum\hours=0 12:\minute\ AM
\else\ifnum\hours<12 \hour:\minute\ AM
\else\ifnum\hours=12 12:\minute\ PM
\else\ifnum\hours>12
\pmhours=\hours
\advance\pmhours by -12
\the\pmhours:\minute\ PM
\fi
\fi
\fi
\fi
}

\def\monthname{\relax\ifcase\month 0/\or January\or February\or
March\or April\or May\or June\or July\or August\or September\or
October\or November\or December\else\number\month/\fi}

\def\bold#1{\setbox0=\hbox{$#1$}     \kern-.025em\copy0\kern-\wd0
\kern.05em\copy0\kern-\wd0
\kern-.025em\raise.0433em\box0 }

\begin{document}

\setcounter{footnote}{1}
\rightline{EFI 24-4}
\rightline{arXiv:2406.05920}
\vskip1.5cm

\centerline{\large \bf 
Possible mixing of a diquark-antidiquark with a $p \bar p$ hadronic molecule}
\bigskip

\centerline{Marek Karliner$^a$\footnote{{\tt marek@tauex.tau.ac.il}}
 and Jonathan L. Rosner$^b$\footnote{{\tt rosner@hep.uchicago.edu}}}
\medskip

\centerline{$^a$ {\it School of Physics and Astronomy}}
\centerline{\it Tel Aviv University, Tel Aviv 69978, Israel}
\medskip

\centerline{$^b$ {\it Enrico Fermi Institute and Department of Physics}}
\centerline{\it University of Chicago, 5640 S. Ellis Avenue, Chicago, IL
60637, USA}
\bigskip
\strut

\begin{center}
ABSTRACT
\end{center}
\begin{quote}
We discuss the possibility that the two nearby resonances observed by
BESIII partially below the \,$p\bar p$\, threshold might be due 
to mixing between two metastable states with the same $J^{PC}=0^{-+}$ 
quantum numbers, but rather different internal structure.  One is a $p \bar p$
hadronic molecule and the other a bound state of a light-quark diquark and an
antidiquark, both with spin 1 and isospin 0,
a composite color antitriplet and triplet, respectively.
The doubling of resonances, one of which may be interpreted as a hadronic
molecule, while the other arises from $q \bar q$ annihilation in a state
with vacuum quantum numbers may be a more general feature
than the specific case considered here.
\end{quote}
\smallskip

\leftline{PACS codes: 12.38.Aw,12.39.Jh,14.40.Rt,36.10.Gv}
\bigskip

\section{Introduction \label{sec:intro}}

Ref.~\cite{BESIII:2023vvr} reports an anomalous line shape of the $X(1840)$
resonance observed in the invariant mass recoiling against the photon in 
the decay $\jpsi\rightarrow\gamma\,3(\pi^+\pi^-)$.
A significant distortion at 1.84 GeV in the line-shape of the
$3(\pi^+\pi^-)$ invariant mass spectrum is observed for the first
time, which could be resolved by two overlapping resonant structures,
$X(1840)$ and $X(1880)$. The new state $X(1880)$ is observed with
a statistical significance larger than $10\sigma$. The mass and
width of $X(1880)$ are determined to be $1882.1\pm1.7\pm0.7$ MeV
and $30.7\pm5.5 \pm2.4$ MeV, respectively, while the mass and width
of the $X(1840)$ are found to be $1842.2^{+7.1}_{-2.6}$ MeV and
$83\pm14\pm11$ MeV, respectively.  The region 
of $\pm15$ MeV around the central value of 1882 MeV extends down to 
1867 MeV, i.e., about 10 MeV below the $p\bar p$ threshold at 1877
MeV. This might indicate the existence of a $p\bar{p}$ bound state.

The history of $p\bar{p}$ bound states goes back at least as far as the
Fermi-Yang interpretation of the pion \cite{Fermi:1949voc}.  Attempts were
made (superseded by the quark model and quantum chromodynamics) to
describe the resonance spectrum \cite{Dalkarov:1979cf} using hadron-antihadron
interactions. Our more modest aim is to explain the two overlapping resonances,
$X(1840)$ and $X(1880)$, in terms of dominant components of their wavefunctions.

Our interpretation may have a broader context, applying to general cases
of two-hadron molecules such as those discussed in Ref.\ \cite{Guo:2017jvc}.
In addition to the states in $3(\pi^+\pi^-)$ there is a broad state $X(1835)$
in the $\eta'\pi^+\pi^-$ channel \cite{ParticleDataGroup:2022pth} with mass
$M(X) = 1826.5^{+15.0}_{-3.4}$ MeV, $\Gamma(X) = 242^{+14}_{-15}$ MeV, whose
participation in the two-state interference we do not consider.  We also
restrict our attention to S-wave proton-antiproton states, ignoring states
with relative angular momentum $\ell > 0$.

\strut\kern-0.5em
We first discuss behavior of wave function components under discrete
symmetries (Sec.~\ref{sec:disc}).  We then (Sec.\ \ref{sec:2st}) interpret
the observed spectrum in terms of two states: a proton-antiproton molecule
and a diquark-antidiquark bound state, whose masses we calculate in
Sec.\ \ref{sec:mass}.  Other examples of similar two-state systems are noted
in Sec.\ \ref{sec:other}, while we conclude in Sec.\ \ref{sec:conc}.

\vrule width 0pt height 2.5ex 
After the first version of this report had been issued,
a parallel set of interpretations
appeared in Ref.~\cite{Ma:2024gsw}.

\section{Quantum numbers} \label{sec:disc}
We begin by identifying the likely $J^{PC}$ quantum numbers of the two
resonances. Both \jpsi\ and the photon have charge-conjugation eigenvalue
{\it C-parity C={-}1}. Therefore both $X(1840)$ and $X(1880)$ must have
$C={+}1$.  The $J^{PC}$ of $X(1840)$ is well-established to be $0^{-+}$
\cite{ParticleDataGroup:2022pth}.  If one interprets $X(1880)$ as a $p \bar p$
resonance, its charge parity is given by
\beq
C_{p\bar p} = ({-}1)^{L+S}
\eeq
where $L$ and $S$ are the orbital and spin angular momentum of the $p \bar p$
system, respectively.  $X(1880)$ is very close to the $p\bar p$ threshold, so
$L$ must be zero, because orbital excitations cost significant energy. We have
$L+S=S=\hbox{even}$. For $p \bar p$ one can have $S=0\
\hbox{or}\ S=1$. It follows that $S=0$ and therefore $J=0$. The parity of a
$p \bar p$ system is $(-1)^{L+1}={-}1$. So finally
\begin{equation}
J^{PC}[X(1840)]= J^{PC}[X(1880)]= 0^{-+}~.
\end{equation}

\section{Two-state interpretation} \label{sec:2st}

If one interprets either $X(1840)$ or $X(1880)$ as a metastable $p \bar p$
bound state, a question arises immediately as to the physical nature of the
other state.  If one of the two states is a hadronic molecule, it is hard to
come up with a plausible scenario in which the other state is also a
hadronic molecule. With only light quarks and antiquarks in the game, it is
hard to generate such a large splitting between below-threshold states.
We are therefore led to search for some completely different configuration
for the state which is not a hadronic molecule, yet has the same quantum
numbers and a similar mass.

\begin{figure}
\begin{center}
\includegraphics[width=0.5\textwidth]{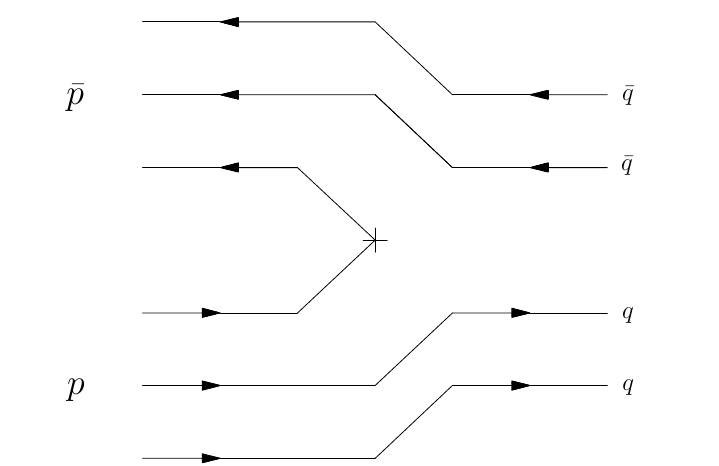}
\end{center}	
\caption{Systems involving contribution of $^3P_0$ amplitude (the symbol +)
to proton--antiproton annihilation into a diquark--antidiquark resonance.
\label{fig:3p0}}	
\end{figure}

The transition to a final state near $\bar p p$ threshold is governed by the
following regularity, illustrated in Fig.~\ref{fig:3p0} \cite{Rosner:1972qde}:
If two hadrons have at least one $q \bar q$ pair in common, they should form at least one resonance via $q \bar q$ annihilation when the center-of-mass
(c.o.m.) 3-momentum is less than some critical value $p^*_0$.
For meson-meson scattering $p_0^{*\kern0.07em M\kern-0.15em M} = 350$ MeV,
while for meson-baryon scattering $p_0^{*\kern0.07em M\kern-0.15em B}=250$ MeV. 
Extrapolating to baryon-antibaryon baryon-antibaryon scattering, one estimates 
$p_0^{*\kern0.07em B\kern-0.15em \bar B} = 200$ MeV \cite{Rosner:1972qde}.
Acceptable intermediate states include any below strong threshold.



Motivated by analogy to the system of the $\chi_{c1}(3872)$ meson [$X(3872)$], 
it has been noted (see, e.g., the discussion in Ref.~\cite{Karliner:2014lta})
that the unusual properties of this meson can be understood if one
assumes that the physical $\chi_{c1}(3872)$ is a mixture of two very
different objects which have the same $J^{PC}=1^{++}$ quantum numbers
and happen to be close in mass: a $\bar D D^*$ hadronic molecule and 
a $P$-wave charmonium, a.k.a. $\chi_{c1}(2P)$.
(One estimate of the $\chi_{c1}(2P)$ mass is 3920.5 MeV \cite{Eichten:2005ga}.)

We now argue that in the $X(1840)$-$X(1880)$ system the analogues are a $p \bar
p$ hadronic molecule and a $P$-wave excitation of a narrow diquark-antidiquark
system.  The possibility of narrow high-mass exotic states has been raised in
\cite{Karliner:2007kf,Karliner:2008rc,Karliner:2021qok}. There it was 
pointed out that narrow high-mass states can arise despite large phase space 
when two nearly degenerate states are coupled to the same dominant decay mode.
A necessary condition for such a proposal is
that the $P$-wave diquark-antidiquark system must have the same quantum numbers
and a mass close to the $p \bar p$ threshold.  Conservation of angular momentum
demands that the quark-antiquark pair which annihilate to leave the diquark-%
antidiquark system must have the quantum numbers $S=1$, $L=1$, and
$J=0$, or $^3P_0$ in a flavor singlet.  The role of {\it production} of a
flavor-singlet$^3P_0$ pair in quarkonium decay has been recognized for many
years \cite{Micu:1968mk,Colglazier:1970vx,Leyaouanc:1972vsx}.
In the following we check the quantum numbers and estimate the mass of the
diquark-antidiquark ground state using the standard toolbox of the 
nonrelativistic quark model.

Let us start from the quantum numbers. We have two spin-1 diquarks, which
are each other's antiparticles: $ud$ (or $uu$) $\bar u\bar d$ 
(or $\bar u\bar u$), obviously with isospin=1, a.k.a. ``bad diquarks". 
Their color representations are $\begin{boldmath}\bar 3_c \end{boldmath}$ and
$\begin{boldmath} 3_c \end{boldmath}$, respectively.
It is clear that they can be combined to a color and isospin singlet.

Next, we have to make sure that we can obtain a bound state with
$J^{PC}=0^{-+}$.
The intrinsic parity of the two-diquark system is ${+}1$. To get
a negative parity, we need to put them in a $P$-wave, i.e., $L=1$. Then to
get the total angular momentum $J=0$, we 
combine the two vectors to $S=1$, so that $\vec J = \vec L + \vec S = 0$.
The consistency check is whether the charge parity comes out right.
For two spin-1 bosons with $S=1$ and $L=1$ we indeed have $C=1$, 
as illustrated by, e.g.,  $\eta(1270)$ with $J^{PC}=0^{-+}$ which 
has a $\rho^+\rho^-$ decay mode \cite{ParticleDataGroup:2022pth}.

\section{Masses} \label{sec:mass}

The calculation of the mass of the diquark-antidiquark system is
straightforward. The mass of each ``bad" diquark is just the constituent
mass of the two light quarks (we ignore isospin breaking), plus
color-hyperfine repulsion in spin-1 state:
\beq
M_{diq,S=1} = 2 m_q + a_{HF} = 2\times 363 + 50 = 776\ \hbox{MeV}.
\eeq
The mass of the diquark-antidiquark state is just the sum of the masses of
the two diquarks, plus the cost of the $P$-wave excitation. The latter is
estimated using a method previously employed in Ref.~\cite{Karliner:2018bms}.
The residual energy difference $\Delta E_R$ after accounting for quark masses
is found to be approximately linear in reduced mass $\mu_R$ of the two-body
system:
\beq \label{eqn:lin}
\Delta E_R = 417.37 - 0.2141 \mu_R
\eeq
The reduced mass of the diquark-antidiquark system is $\mu_R = (1/2)(776) =
388$ MeV, so $\Delta E_R = 334$ MeV and we estimate the mass of the
diquark-antidiquark configuration with $J^{PC}=0^{-+}$:
\beq \label{eqn:est}
M_{diq-diq} = 2 \times 776 + 334 = 1886 \ \hbox{MeV}.
\eeq

In order for the molecular proton-antiproton state to be near the threshold
of $2m_p = 2(938.27) = 1876.5$ MeV, we must identify it with the $X(1880)$,
leaving the $X(1840)$ to be the state predicted in Eq.\ (\ref{eqn:est}).
This is probably acceptable, given the crude nature of the approximation
(\ref{eqn:lin}).  The opposite order, with $X(1840)$ identified as the $p \bar
p$ molecule and $X(1880)$ taken as the diquark-antidiquark candidate,
implies an unexpectedly large molecular binding energy.

\section{Other systems} \label{sec:other}

The doubling of resonances, one of which may be interpreted as a hadronic
molecule while the other arises from $q \bar q$ annihilation in a state
with vacuum quantum numbers, i.e., $^3P_0$, may be a more general feature
than the case considered here.  One would then expect such resonances
as $f_0(980)$, $\Lambda(1405)$, and $D_s(2317)$ (see Table I of Ref.\
\cite{Guo:2017jvc}) to be accompanied by a partner at a nearby mass.

\subsection{$f_0(980)$}

The $f_0(980)$ is a spin-zero isospin-zero resonance with couplings to
$\pi \pi$, $K \bar K$, $u \bar u$, $d \bar d$, $s \bar s$, $\eta \pi \pi$,
nucleon-antinucleon, $\gamma \gamma$, and tetraquark configurations.  These
final states are sorted out in an extensive section of the Particle Data
Group's Review of Particle Physics \cite{Scalars:2023}.  Analogy with the
doubled-resonance picture would regard it as a combination of a $K \bar K$
molecule with small binding energy and a quark-antiquark $^3P_0$ state with 
a mixture of $^3P_0$ configurations $(u\bar u + d\bar d)/\sqrt2$ and $s\bar s$.
The doubled-resonance structure of the $f_0(980)$ is consistent with the
analysis of Ref.\ \cite{Au:1986mq}.  However, recent reviews of the status of
the $f_0(980)$ region (\cite{Yao:2020bxx,Garcia-Martin:2011nna}) need only one
pole.  The former summarizes: ``...the $f_0(980)$ is close to the $\bar K^0K^0$
threshold (995 MeV) and the width is near 40 MeV.  There is always only one
pole in ... the $\pi \pi \to K K$ coupled channels and it is most 
likely to be a $KK$ molecule.'' The latter finds a $f_0(980)$ pole at
$996 \pm 7 -i~25^{+10}_{-6}$ MeV.

\subsection{$\Lambda(1405)$}

The $\Lambda(1405)$ is one of the first hadron resonances, making its appearance
before 1960 \cite{Dalitz:1960}.  It couples to the open channel $\pi \Sigma$
(threshold $\sim 137 + 1193 = 1330$ MeV) and the closed channel $\bar K N$
(threshold $\sim 495 + 939 = 1434$ MeV).  It is represented by a two-pole
structure cite{Hyodo:2021}, with pole 1 very close to the $\bar KN$ threshold
and imaginary part corresponding to a rather narrow resonance.  There is less
agreement about the position of pole 2 but its real part is below 1400 MeV
and its imaginary part corresponds to a broader resonance than pole 1.

Proceeding by analogy with the system $X(1840-1880)$, we would identify pole 1
(narrow, near threshold) with $X(1880)$, and pole 2 (broader, some distance
below threshold) with $X(1840)$.

\subsection{$D_{s0}(2317)$}

A relativistic quark model of charmed nonstrange and strange meson masses
\cite{Ebert} predicts all $S$-wave and $P$-wave masses to a satisfactory
extent with the exception of the spin-zero $D_{s0}(2317)$ and spin-1
$D_{s1}(2460)$.  Its prediction of $M(D_{s0}~(^3P_0))$ is compared with others
in Table III of Ref.\ \cite{Ni}, summarized in Table 1.

%
\begin{table}[h]
\caption{Predictions of $M(D_{s0}~(^3P_0))$ (\cite{Ni})(MeV)
\label{tab:comp}}
\begin{center}
\begin{tabular}{c c c c c c} \\ \hline
   2409    &   2484     &  2509     &  2380   &   2344    &    2409 \\
\cite{Ni}&\cite{Godfrey}&\cite{Ebert}&\cite{Zeng}&\cite{Li}&\cite{Lahde}\\
\hline
\end{tabular} 
\end{center}
\end{table} 

Ref.\ \cite{Ebert} finds the relativistic $D_{s0}$ mass higher than any other
calculation, and farther from the observed value of $D_{s0}$.  We propose
that $D_{s0}(2317)$ be accepted as the analog of $X(1840)$ while there should
exist a distinct molecular state near $D^0 K^+$ threshold (1864.8 + 493.7 =
2359 MeV) which would be the analog of $X(1880)$.

\section{Conclusions} \label{sec:conc}

Our treatment of the doubled resonance around 1840 and 1880 MeV has several
features:

\vrule width 0pt height 3.5ex 
(a) The $J^{PC}$ quantum numbers work out for both proton-antiproton (molecule,
narrow, X(1880), near threshold) and diquark-antidiquark (broader, X(1840)).

(b) The masses work out almost by construction for proton-antiproton and rather
nontrivially for diquark-antidiquark.

(c) We have identified a simple and non-trivial mechanism allowing the
transition between the two configurations.

\vrule width 0pt height 3.0ex 
A consequence of our conjecture is that the
spin-parity of the $3(\pi^+\pi^-)$ system must be determined to be $0^{-+}$
without interference from the broad $\eta'\pi^+\pi^-$ state or others
around 1835 MeV.

\section*{Acknowledgements}
We thank Changzheng Yuan and Shuangshi Fang for useful discussions.
This research was supported in part by the ISF-NSFC joint research program
(grant No. 3166/23).


\begin{thebibliography}{99}

\bibitem{BESIII:2023vvr} M.~Ablikim \textit{et al.} [BESIII Coll.],
{\em ``Observation of the Anomalous Shape of $X(1840)$ 
in $J/\psi\rightarrow\gamma3(\pi^+\pi^-)$ Indicating a Second Resonance 
Near $p\bar p$ Threshold"},
Phys.\ Rev.\ Lett.\ \textbf{132}, 151901 (2024) [[arXiv:2310.17937 [hep-ex]].

\bibitem{Fermi:1949voc} E.~Fermi and C.~N.~Yang,
{\em Are mesons elementary particles?}` Phys.\ Rev.\ \textbf{76}, 1739 (1949).

\bibitem{Guo:2017jvc} F.~K.~Guo, C.~Hanhart, U.~G.~Mei\ss{}ner, Q.~Wang,
Q.~Zhao and B.~S.~Zou,{\em ``Hadronic molecules,''}
Rev.\ Mod.\ Phys.\ \textbf{90}, 015004 (2018)
[erratum: Rev.\ Mod.\ Phys.\ \textbf{94}, 029901 (2022)]
[arXiv:1705.00141 [hep-ph]].

\bibitem{Ma:2024gsw} B.~Q.~Ma,
{\em``Protonium: Discovery and Prediction,''} [arXiv:2406.19180 [hep-ph]].

\bibitem{ParticleDataGroup:2022pth}
R.~L.~Workman \textit{et al.} [Particle Data Group],
{\em ``Review of Particle Physics,''}
PTEP \textbf{2022}, 083C01 (2022).

\bibitem{Dalkarov:1979cf} O.~D.~Dalkarov, V. B. Mandelzweig, and I.  S.
Shapiro, {\em On Possible Quasinuclear Nature of Heavy Meson Resonances,"}
Nucl.\ Phys.\ {\bf A21}, 88 (1970);  O. D. Dalkarov, {\em ``Quiasinuclear
nuclear states in baryon-antibaryon systems,"}  1980.

\bibitem{Rosner:1972qde} J. L. Rosner, {\em ``Compulsory Resonance Formation,"}
Phys.\ Rev.\ D {\bf 7}, 2717 (1972).

\bibitem{Karliner:2014lta} M.~Karliner and J.~L.~Rosner,
{\em ``$X(3872)$, $X_b$, and the $\chi_{b1}(3P)$ state"},
Phys.\ Rev.\ D \textbf{91}, 014014 (2015) [arXiv:1410.7729 [hep-ph]].

\bibitem{Eichten:2005ga}E.~J.~Eichten, K.~Lane and C.~Quigg, {\em ``New states
above charm threshold,''}Phys.\ Rev.\ D \textbf{73}, 014014 (2006)
[erratum: Phys.\ Rev.\ D \textbf{73}, 079903 (2006)]
[arXiv:hep-ph/0511179 [hep-ph]].

\bibitem{Karliner:2007kf} M.~Karliner and H.~J.~Lipkin,
{\em ``Possibility of Narrow High-Mass Exotic States,''}
[arXiv:0710.4055 [hep-ph].

\bibitem{Karliner:2008rc} M.~Karliner and H.~J.~Lipkin,
{\em``Possibility of Exotic States in the Upsilon system,''}
[arXiv:0802.0649 [hep-ph]].

\bibitem{Karliner:2021qok} M.~Karliner and J.~L.~Rosner,
{\em``Configuration mixing in strange tetraquarks Zcs,''}
Phys.\ Rev.\ D \textbf{104}, 034033 (2021) [arXiv:2107.04915 [hep-ph]].

\bibitem{Micu:1968mk} L.~Micu,
{\em ``Decay rates of meson resonances in a quark model,''}
Nucl.\ Phys.\ B \textbf{10}, 521 (1969).

\bibitem{Colglazier:1970vx} E.~W.~Colglazier and J.~L.~Rosner,
{\em ``Quark graphs and angular distributions in positive parity meson
decays,''} Nucl.\ Phys.\ B \textbf{27}, 349 (1971).

\bibitem{Leyaouanc:1972vsx} A.~Le Yaouanc, L.~Oliver, O.~Pene and J.~C.~Raynal,
{\em``Naive quark pair creation model of strong interaction vertices,''}
Phys.\ Rev.\ D \textbf{8}, 2223 (1973).

\bibitem{Karliner:2018bms} M.~Karliner and J.~L.~Rosner,
{\em ``Scaling of $P$-wave excitation energies in heavy-quark systems,''}
Phys.\ Rev.\ D \textbf{98}, 074026 (2018), [arXiv:1808.07869 [hep-ph]].

\bibitem{Scalars:2023} T.~Gutsche, C.~Hanhart, R.~E.~Mitchell, and S.~Spanier,
{\em ``Scalar Mesons below 1 GeV,''} Section 64 in 
\cite{ParticleDataGroup:2022pth}.

\bibitem{Au:1986mq} K.~L.~Au, M.~R.~Pennington and D.~Morgan,
{\em ``Experimental Evidence for Dynamics Beyond the Quark Model: An Extra I=0
Scalar Meson Near 1-{GeV},''} Phys.\ Lett.\ B \textbf{167}, 229 (1986).

\bibitem{Yao:2020bxx} D.~L.~Yao, L.~Y.~Dai, H.~Q.~Zheng and Z.~Y.~Zhou,
{\em``A review on partial-wave dynamics with chiral effective field theory and
dispersion relation,'' Rept.\ Prog.\ Phys.\ \textbf{84}, 076201 (2021)
[arXiv:2009.13495 [hep-ph]].

\bibitem{Garcia-Martin:2011nna} R. Garcia-Martin, R. Kaminski, J. R. Pelaez,
and J. Ruiz de Elvira, Phys.\ Rev.\ Lett.\ {\bf 107}, 072001 (2011)
[arXiv:1107.1635 [hep-ph]].

\bibitem{Dalitz:1960} R. H. Dalitz and S. F. Tuan, Phys.\ Rev.\ Lett.\
{\bf 2}, 425 (1959); Ann.\ Phys.\ N. Y.  {\bf 10}, 307 (1960).

\bibitem{Hyodo:2021} T. Hyodo and U. G. Meissner, {\em`` Pole Structure of
the $\Lambda(1405)$ Region,''} S. Navas \textit{et al.} [Particle Data Group],
{\em ``Review of Particle Physics,''} Phys.\ Rev.\ D \textbf 110}, 030001
(2024), Chapter 83.

\bibitem{Ebert} D.~Ebert, R.~N.~Faustov and V.~O.~Galkin,
{\em ``Heavy-light meson spectroscopy and Regge trajectories in the
relativistic quark model,''} Eur.\ Phys.\ J. C \textbf{66},
197 (2010) [arXiv:0910.5612 [hep-ph]].

\bibitem{Ni} R.~H.~Ni, Q.~Li and X.~H.~Zhong,
{\em ``Mass spectra and strong decays of charmed and charmed-strange mesons,''}
Phys.\ Rev.\ D \textbf{105}, 056006 (2022) [arXiv:2110.05024 [hep-ph]].

\bibitem{Godfrey} S.~Godfrey and K.~Moats,
{\em ``Properties of Excited Charm and Charm-Strange Mesons,''}
Phys.\ Rev.\ D \textbf{93}, 034035 (2016) [arXiv:1510.08305 [hep-ph]].

\bibitem{Zeng} J.~Zeng, J.~W.~Van Orden and W.~Roberts,
{\em ``Heavy mesons in a relativistic model,''}
Phys.\ Rev.\ D \textbf{52}, 5229 (1995) [arXiv:hep-ph/9412269 [hep-ph]].

\bibitem{Li} D.~M.~Li, P.~F.~Ji and B.~Ma,
{\em ``The newly observed open-charm states in quark model,''}
Eur.\ Phys.\ J. C \textbf{71}, 1582 (2011) [arXiv:1011.1548 [hep-ph]].

\bibitem{Lahde} T.~A.~Lahde, C.~J.~Nyfalt and D.~O.~Riska,
{\em ``Spectra and M1 decay widths of heavy light mesons,''}
Nucl.\ Phys.\ A \textbf{674}, 141 (2000) [arXiv:hep-ph/9908485 [hep-ph]].

\end{thebibliography}
\end{document}